\documentclass[conference]{IEEEtran}
\IEEEoverridecommandlockouts
\usepackage{cite}
\usepackage{amsmath,amssymb,amsfonts}
\usepackage{algorithmic,multirow}
\usepackage{array}
\usepackage{graphicx}
\usepackage{gensymb}
\usepackage{textcomp}
\usepackage{xcolor}
\usepackage{color}
\usepackage{array}

\def\BibTeX{{\rm B\kern-.05em{\sc i\kern-.025em b}\kern-.08em
    T\kern-.1667em\lower.7ex\hbox{E}\kern-.125emX}}
\begin{document}

\title{Unsupervised Adversarial Domain Adaptation for Cross-Lingual Speech Emotion Recognition}

\author{\IEEEauthorblockN{Siddique Latif}
\IEEEauthorblockA{\textit{University of Southern Queensland} \\
Australia\\
siddique.latif@usq.edu.au}
\and
\IEEEauthorblockN{Junaid Qadir}
\IEEEauthorblockA{\textit{Information Technology University (ITU)} \\
Lahore, Pakistan. \\
junaid.qadir@itu.edu.pk}
\and
\IEEEauthorblockN{Muhammad Bilal}
\IEEEauthorblockA{\textit{University of the West of England (UWE)} \\
Bristol, United Kingdom. \\
muhammad.bilal@uwe.ac.uk}

 }

\maketitle

\begin{abstract}
Cross-lingual speech emotion recognition (SER) is a crucial task for many real-world applications. The performance of SER systems is often degraded by the differences in the distributions of training and test data. These differences become more apparent when training and test data belong to different languages, which cause a significant performance gap between the validation and test scores. It is imperative to build more robust models that can fit in practical applications of SER systems. Therefore, in this paper, we propose a Generative Adversarial Network (GAN)-based  model for multilingual SER. Our choice of using GAN is motivated by their great success in learning the underlying data distribution.  The proposed model is designed in such a way that can learn language invariant representations without requiring target-language data labels. We evaluate our proposed model on four different language emotional datasets, including an Urdu-language dataset to also incorporate alternative languages for which labelled data is difficult to find and which have not been studied much by the mainstream community. Our results show that our proposed model can significantly improve the baseline cross-lingual SER performance for all the considered datasets including the non-mainstream Urdu language data without requiring any labels. 

\end{abstract}


\section{Introduction}
Speech emotion recognition (SER) is gaining more interest in recent years. The goal of SER is to identify different kinds of human emotion from the given speech, which has been proven very helpful in automating many real-life applications including health-related diagnostics \cite{latif2017mobile,latif20175g,rana2019automated,latif2018mobile}. Existing SER systems can perform to a satisfactory level when training and test data belong to the same corpus \cite{Latif2018v,latif2019direct}. However, it is still an open challenge to design more robust SER systems that are more resilient to cross-lingual emotions recognition. 

Due to recent advancement in the field of machine learning (ML), particularly deep learning (DL), researchers are attempting to solve various problems in audio and related fields \cite{wei2012distributed,rana2016gait,latif2018phonocardiographic,qayyum2018quran}. For SER. many researchers are also attempting to design more robust systems that can work best with applications involving multiple languages. One approach for the design of more robust SER systems is to use as diverse dataset (having multilingual data) as possible. Studies (e.g., \cite{schuller2010cross,latif2018cross}) have shown that a model trained using multiple sources or corpora can help to achieve better results for SER. However, acoustic training from multiple language data is not a reasonable approach as it requires labelled data that might not be available for all languages. Alternatively, robustness in SER system can also be achieved by using a partial data from the source language to improve the performance \cite{Latif2018,zhou2019transferable}; but some labelled target data for training the model is needed here as well. 

A more practical approach is the use of \textit{domain adaptation}, which generalises SER systems to the multilingual scenarios without the need for labelled data. Researchers have tried different domain adaptation methods in SER to improve the performance of models on cross-lingual or cross-corpus emotion recognition tasks \cite{sagha2016cross,deng2017universum}. To this end, unsupervised domain adaptation methods are becoming very popular in such applications. Recently, Generative Adversarial Networks (GANs) \cite{goodfellow2014generative} has become very popular and being employed in various vision \cite{latif2018automating} and speech related fields \cite{latif2018adversarial}. They are also exploited for unsupervised domain adaptation in several tasks related to voice user applications including speaker identification \cite{wang2018unsupervised},  automatic speech recognition (ASR) \cite{shinohara2016adversarial,sun2017unsupervised,meng2018speaker} and SER \cite{abdelwahab2018domain,tu2019towards}. However, multilingual SER is not explored using GAN based domain adaptation approaches. 

In this study, we propose an adversarial domain adaptation for multilingual SER, particularly for languages like Urdu for which emotional labels are not available.  Urdu is the official national language of Pakistan and is amongst the 22 official languages recognised in the Constitution of India. The performance of state-of-the-art SER systems degrades when unknown language data like Urdu is used in testing phase \cite{albornoz2017emotion}. To the challenges introduced by such languages, it is crucial for developing more robust SER systems for their next-generation cross-cultural applications. Therefore, we evaluate the proposed approach on multilingual SER tasks including Urdu language data.

We assume that source language data with annotated emotional labels is available and is used for training the model while the target language data (which is considered as the target domain and is used for testing the model's performance) does not have emotional labels. The proposed approach uses unsupervised adversarial domain adaptation for multilingual SER where the model aims to learn language invariant emotional representation from the given source language similar to target language features. 

\begin{figure}[!ht]
\centering
\includegraphics[width=0.4\textwidth]{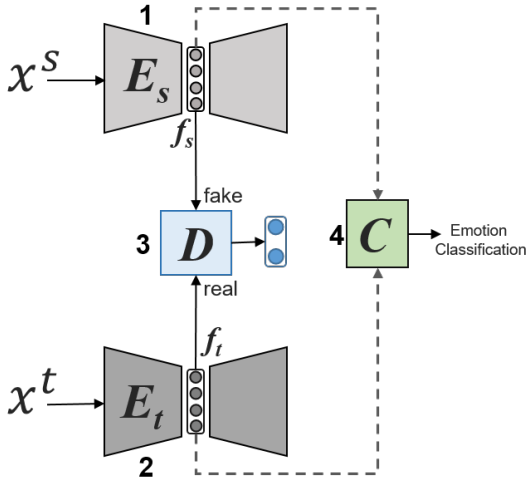}
\caption{Proposed model for multilingual domain adaptation. Where $E_{s}$ and $E_{t}$ are source language and target language encoders, $D$ is the discriminator, and $C$ is the classifier}
\label{fig:Model}
\end{figure}

Our proposed model utilizes the following four networks in its architecture (illustrated in Figure \ref{fig:Model}): 
\begin{itemize}
    \item (1 and 2) feature encoding networks for the source language data and target language data, respectively; 
    \item (3) a discriminator network that discriminates between the features encoded by source encoding network and by target encoding network; and 
    \item (4) a classifier for emotion identification.
\end{itemize}

 We have evaluated the proposed model on four publicly available language datasets and have compared our results with popular emotional features eGeMAPS. Our results indicate that our approach significantly improves the multilingual emotion identification predictive accuracy. Our results are a promising advance in the field since our model can be used to identify emotions for any unlabelled language data in an unsupervised adversarial manner.  

\section{Related Work}
Cross-language SER is important and has been studied in various research works. It aims to generalise the classifiers having different training and test conditions including different noise levels, microphone settings, and speaker variations and language. The performance of classifier on these different conditions have been highlighted in \cite{elbarougy2014toward,schuller2010cross,eyben2010cross,parlak2014cross}. These studies used SVM as a classifier and pointed out the need for in-depth research to improve the performance of cross-lingual SER. Researchers have also attempted different classifiers to improve the performance of cross-lingual SER. For instance, Neumann et al. \cite{neumann2018cross} used CNN for binary arousal/valence classification for the French and English languages. They showed that fine-tuning the model on the target domain can help to produce better results. In another work, Albornoz et al. \cite{albornoz2017emotion} developed an emotion profile-based ensemble SVM SER for different languages and demonstrated a substantial performance gain over baseline accuracy by training the ensemble model in a language-agnostic manner. Likewise, Li et al. \cite{li2019improving} used a combination of different features along with speaker normalisation technique to improve multi-lingual emotion recognition. 

Previous works have also attempted different techniques to minimise the differences in the feature space of both source and target domains. Zhang et al. \cite{zhang2011unsupervised} normalised the features of each corpus separately to decrease cross-corpus variability. To reduce the effect of covariate shift, Hassan et al. \cite{6488742} used three algorithms to apply importance weights to the training data of SVM classifier to match the test data distribution. Deng et al. \cite{deng2014autoencoder} used autoencoders with shared hidden layer to learn common feature representation across different datasets. These authors were able to reduce the discrepancy among corpora and improve the baseline results. In another study \cite{7862157}, authors used Universum autoencoder, to enhance the performance of SER in mismatched training and test conditions. They achieved promising results on cross-corpus evaluations due to the additional supervised learning capability of Universum autoencoders in contrast to conventional autoencoders. 

In contrast to the studies mentioned above, we aim to minimise mismatch of training and test conditions in an unsupervised adversarial way for cross-lingual emotion recognition. Our attempt is motivated by the great success of GANs in computer vision for domain adaptation task \cite{bousmalis2017unsupervised,tzeng2017adversarial}. Very few studies exploited GANs for SER.  For instance, Zhou et al. \cite{zhou2018transferable} used a class-wise domain adaptation method using adversarial training to address cross-corpus mismatch issue. They used two datasets, including AIBO  and EMO-DB for the same language and showed that adversarial training is useful when the model is to be trained on target language with minimal labels. Wang et al. \cite{wang2018unsupervised} exploited adversarial multitask training to learn a common representation for both the source and target language domains. Authors used two datasets of the English language and presented promising results for cross-corpus emotion recognition by creating more discriminant features that reduce the gap between the source and target datasets. Similarly, Gideon et al. \cite{gideon2019barking} used three well-known English language datasets and proposed an adversarial discriminative domain generalisation method for cross-corpus emotion recognition. Besides these studies, we proposed a method for cross-lingual emotion recognition over four different language datasets. The proposed model is then evaluated for languages such as Urdu whose labelled emotional data is barely available.

\begin{table*}[!ht]
\centering
\caption{Selected corpora information and the mapping of class labels onto Negative/Positive valence.}
\begin{tabular}{|m{1.2cm}|m{1.3cm}|m{1cm}|m{1.3cm}|m{4.7cm}|m{3.28cm}|m{1.4cm}|}
\hline
\textbf{Corpus}
&\textbf{Language}
&\textbf{Age}
&\textbf{Utterances}
&\textbf{Negative Valance}
&\textbf{Positive Valance}
&\textbf{References}
\\ \hline

 \begin{tabular}[c]{@{}l@{}}EMO-DB\end{tabular}
&\begin{tabular}[c]{@{}l@{}}German\end{tabular}
&\begin{tabular}[c]{@{}l@{}}Adults\end{tabular}
&\begin{tabular}[c]{@{}l@{}}494\end{tabular}
&\begin{tabular}[c]{@{}l@{}}Anger, Sadness, Fear, Disgust, Boredom\end{tabular} 
&\begin{tabular}[c]{@{}l@{}}Neutral, Happiness\end{tabular}
&\begin{tabular}[c]{@{}l@{}}\cite{burkhardt2005database}\end{tabular}
\\ \hline
 \begin{tabular}[c]{@{}l@{}}SAVEE\end{tabular}
&\begin{tabular}[c]{@{}l@{}}English\end{tabular}
&\begin{tabular}[c]{@{}l@{}}Adults\end{tabular}
&\begin{tabular}[c]{@{}l@{}}480\end{tabular}
&\begin{tabular}[c]{@{}l@{}}Anger, Sadness, Fear, Disgust \end{tabular}
&\begin{tabular}[c]{@{}l@{}}Neutral, Happiness, Surprise\end{tabular}
&\begin{tabular}[c]{@{}l@{}}\cite{jackson2014surrey}\end{tabular}
\\ \hline 
 \begin{tabular}[c]{@{}l@{}}EMOVO\end{tabular}
&\begin{tabular}[c]{@{}l@{}}Italian\end{tabular}
&\begin{tabular}[c]{@{}l@{}}Adults\end{tabular}
&\begin{tabular}[c]{@{}l@{}}588\end{tabular}
&\begin{tabular}[c]{@{}l@{}}Anger, Sadness, Fear, Disgust\end{tabular}
&\begin{tabular}[c]{@{}l@{}}Neutral, Joy, Surprise\end{tabular}
&\begin{tabular}[c]{@{}l@{}}\cite{costantini2014emovo}\end{tabular}
\\ \hline
 \begin{tabular}[c]{@{}l@{}}URDU\end{tabular}
&\begin{tabular}[c]{@{}l@{}}Urdu\end{tabular}
&\begin{tabular}[c]{@{}l@{}}Adults\end{tabular}
&\begin{tabular}[c]{@{}l@{}}400\end{tabular}
&\begin{tabular}[c]{@{}l@{}}Angry, Sad\end{tabular}
&\begin{tabular}[c]{@{}l@{}}Neutral, Happy\end{tabular}
&\begin{tabular}[c]{@{}l@{}}\cite{latif2018crossl}\end{tabular}
\\ \hline

\end{tabular}
\centering
\label{table: MAP}
\end{table*}
\section{Model}
We proposed a model for unsupervised domain adaptation of multilingual SER. Our proposed model leverages the unlabelled target language data and aims to learn common feature representations for both source and target language. 

Fig. \ref{fig:Model} shows the architecture of proposed model. Where source language data $x^{s}$ is fed to source language feature encoder $E_{s}$ and target language data $x^{t}$ is given to the target language feature encoder $E_{t}$. Both $E_{s}$ and $E_{t}$ are connected to discriminator $D$ that is tasked to enforce $E_{s}$ to learn features $f_{s}$ from $x^{s}$ which are similar to $f_{t}$ of target language data $x^{t}$. The intuition of this method is that emotional data of different languages have some common features \cite{Latif2018,latif2018crossl,elbarougy2014toward,schuller2010cross} that we are learning in this model in an unsupervised way. The features ($f_{s}$ and  $f_{t}$ ) encoded by $E_{s}$ and $E_{t}$ are given to the classifier $C$ for classification. Here we trained $C$ on $f_{s}$ by considering that source language data has labels while test is performed on target language data without considering its labels. 

Our proposed model is trained like GANs \cite{goodfellow2014generative} via an adversarial process to produce features $f_{s}$ similar to the target domain. In a simple GAN, the generator $G_{s}$ maps the latent vectors $z$ drawn from some known prior $p_{z}$ ( e.g. Gaussian) to fake data points $G(z)$. The discriminator $D _{s}$ is tasked with differentiating between samples generated $G(z)$ (fake) and real data samples $x$ (drawn from a distribution $p_{\text{data}}$). Both generator $G_{s}$ and a discriminator $D_{s}$ play two-player min-max game using the following GAN loss:
\begin{multline}
\label{Gan}
    \underset{G_{s}}{\text{min}} \  \underset{D_{s}}{\text{max}} V(D_{s},G_{s})=\mathrm{E}_{x \sim p_{\text{data}}}[\log(D_{s}(x))]\\ + \mathrm{E}_{z \sim p_{z}}[\log(1 - D_{s}(G_{s}(z)))]
\end{multline}
Here, we use GAN, where instead of latent vectors $z$, generator $E_{s}$ is fed by source language features $x^{s}$ and is trained to learn representations $f_{s}$ similar to target language representations $f_{t}$ encoded by $E_{t}$. The adversarial loss training loss $\mathcal{L}_{\text{adv}}$ for $E_{s}$ is defined as:
\begin{equation}
\label{adv}
    \mathcal{L}_{\text{adv}}  = \log( 1 -  D(E_{s}(x^s)))
\end{equation}
The generator $E_{s}$ attempt to fool discriminator $D$ by generating features very similar to $f_{t}$. The discriminator $D$ classifies whether features are drawn from the source language ($f_s$) as fake or the target language ($f_{t}$) as real using the following loss function:
\begin{equation}
    \mathcal{L}_{\text{dic}}  = -\log\big(\sum_{i=0}^{N_{s}}D(E_{s}(x_{i}^{s})) -  \sum_{j=0}^{N_{s}}(1-D(E_{t}(x_{j}^{t}))\big)
\end{equation}

where $N_{s}$ and $N_{t}$ are the training samples for source and target language data respectively. In our model, both $E_{s}$ and $E_{t}$ are two autoencoder networks that encode the source and target data in latent code. The intuition of using autoencoders is that they encode the given data into underlying feature structures using reconstruction loss \cite{deng2013sparse,usman2017using}.  Both autoencoders were trained separately using reconstruction loss while the encoder part $E_{s}$ of source autoencoders is updated using the adversarial loss (see Equation \ref{adv}) to learn language invariant features. We use an SVM as the classifier $C$ for emotion identification.   

\section{Experimental Setup}

We have selected EMO-DB, SAVEE, EMOVO, and URDU datasets in this work. The selection of these corpora is made to incorporate maximum diversity of languages especially to cover infrequently analysed languages such as Urdu. Further details on selected databases, speech features, and model configuration are presented below.

\subsection{Speech Databases}
\subsubsection{EMO-DB}
It is a well known and widely used corpus for SER. The language of EMO-DB dataset is German, and it was introduced by \cite{burkhardt2005database}. It comprises the recordings of ten professional actors in 7 emotions: anger, disgust, boredom, fear, neutrality, joy, and sadness. The linguistic content used for recordings is pre-defined emotionally neutral ten short sentences in the German language. Overall, it contains over 700 utterances, while only 494 utterances are emotionally labelled. We used only annotated utterances in this work.

\subsubsection{SAVEE}
Surrey Audio-Visual Expressed Emotion (SAVEE) database \cite{jackson2014surrey} is another popular multimodal emotional dataset. It includes the recordings from four male actors in 7 different emotions. The language of SAVEE dataset is British English. The recordings in this dataset were evaluated by ten different assessors under visual, audio,  and audio-visual conditions to assure the quality of emotional acting. The scripts used for data recordings were selected from the standard TIMIT corpus \cite{garofolo1993darpa}. In total, SAVEE contains 480 utterances in 7 emotions: neutral, happiness, sadness, anger, surprise, fear, and disgust. We used all these emotions in our experiments by mapping them on the binary valance.

\subsubsection{EMOVO}
This dataset is the first Italian language emotional corpus and contains 588 recordings \cite{costantini2014emovo}. There are 6 actors whose scripts of 14 different sentences in 7 different emotional states including disgust, fear, anger, joy, surprise, sadness and neutral. The recordings were evaluated by two separate groups of listeners to validate the performance of emotional actors. All the recordings in this corpus were made with equipment in the Fondazione Ugo Bordoni laboratories. 

\subsubsection{URDU}
This corpus is the first Urdu language dataset that includes unscripted and spontaneous emotional speech \cite{latif2018cross}. It comprises the audio recordings collected from the discussion of the different guests of Urdu TV talk shows. In total, 400 utterances for four basic emotions (angry, happy, sad, and neutral) were collected. The recordings were given to four different annotators who were tasked to annotate them based on audio-visual conditions. There are 38 speakers including 27 males and 11 females. We utilised all 400 utterances in this work.



\subsection{Feature Extraction}

In this study, we have used a minimalistic feature set called eGeMAPS \cite{eyben2016geneva}. These features are widely used frame-level knowledge-inspired parameters. The eGeMAPS comprise Low-Level Descriptor (LLD) of speech which has been suggested as the most descriptive emotional feature by paralinguistic studies. Besides, eGeMAPS features also provide performance comparable and even better compared to large brute-force features\cite{eyben2016geneva}. In total, eGeMAPS consists of 88 parameters related to spectral, energy, frequency, cepstral, and dynamic information. The components of eGeMAPS selected from the arithmetic mean and coefficient of variation of 18 LLDs, 6 temporal features, 8 functionals applied to loudness and pitch, 4 statistics over the unvoiced segments, and 26 additional dynamic parameters and cepstral parameters. A list of these LLDs and functionals can be found in Section 3 of \cite{eyben2016geneva}. We computed eGeMAPS using openSMILE toolkit \cite{eyben2010opensmile}.

\subsection{Model Configuration}
We implemented our model using the Tensorflow library. Both the encoder parts ($E_s$ and $E_t$) consist of two fully connected (FC) layers with the latent code dimension of $512$. The discriminator $D$ also consists of two FC layers having 512 and 256 hidden units followed by a softmax layer. For regularisation, we used dropout layer between FC layers of $E_s$, $E_t$ and $D$, with a dropout rate of $0.5$. We trained the models using the training set, and the validation set was used for hyper-parameter selection.  For minimisation of cross-entropy loss function of discriminator $D$, we used RMSProp optimiser \cite{tieleman2012lecture}, with an initial learning rate of $10^{-4}$. We input speech segments of length $250$ms into the model for encoding the latent code of dimension $512$. The selection of speech segment is made based on the previous studies \cite{provost2013identifying,wollmer2013lstm}. The latent code encoded by both $E_s$ and $E_t$ are then fed to SVM. The utterance level prediction, in testing phase, is obtained by averaging the posterior probabilities of the respective segments. For all experiments, the validation is performed within corpus in a speaker-independent manner to pick the optimal hyper-parameters. We select an RBF kernel due to its better performance compared to the linear and cubic kernel during experimentation.

\section{Experiments and Results}
This section reports the experimental evaluations of the proposed model for cross-lingual SER. For this, we used four publicly available datasets. These databases are annotated differently; therefore, we consider binary positive/negative valence classification problem in this study (see Table \ref{table: MAP}). We adopt the binary valence mapping of categorical emotions from \cite{5557843, Latif2018}. The input features of audio utterances are given to the model for encoding them into language invariant representations which are passed to SVM for classification. We performed experiments in speaker independent evaluation scheme for all datasets and results are reported in terms of unweighted average recall rate (UAR).  For the Urdu dataset, we used 30 speakers as training data and the remaining 8 for testing with five-fold cross-validation. For other corpora, we used one-speaker-out evaluation scheme with cross-validation equal to the number of speakers in the respective dataset as per the accepted practice for computing the baseline results of SER \cite{eyben2016geneva}. 

\begin{table}[!ht]
\centering
\caption{Baseline results within corpus using eGeMAPS}
\begin{tabular}{|c|c|c|c|c|}
\hline
Corpus   & EMO-DB & SAVEE & EMOVO & URDU  \\ \hline
UAR (\%) & 81.3  & 65.1 & 74.2 & 83.4 \\ \hline
\end{tabular}
\label{table: baseline}
\end{table}
For baseline results, we trained SVMs using eGeMAPS features to perform classification within a corpus using both training and testing data from the same language. The obtained baseline results provide us with an idea about the best achievable accuracy within each corpus. Table \ref{table: baseline} shows the baseline results for all datasets.

\begin{table}[!ht]
\scriptsize
\centering
\caption{UAR (\%) comparison cross-lingual emotion recognition using latent code learnt by proposed models with eGeMAPS}
\begin{tabular}{|c|c|c|c|c|}
\hline
\multirow{2}{*}{Source} & \multirow{2}{*}{Target} & \multicolumn{3}{c|}{UAR (\%)}               \\ \cline{3-5} 
                        &                         & eGeMAPS & Latent Codes & eGeMaps+Latent Codes \\ \hline
EMO-DB                   & \multirow{3}{*}{URDU}   & 57.8        & 64.2           &      65.2              \\ \cline{1-1} \cline{3-5} 
SAVEE                   &                         & 45.8        & 52.3          &      58.0                \\ \cline{1-1} \cline{3-5} 
EMOVO                   &                         &  40.1       & 50.5         &    53.6               \\ \hline
\multirow{3}{*}{URDU}   & EMODB                   & 55.1        & 64.5          &     65.3              \\ \cline{2-5} 
                        & SAVEE                   & 43.7        &51.8           &      53.2               \\ \cline{2-5} 
                        & EMOVO                   & 50.8        & 59.8          &     61.3               \\ \hline
\end{tabular}
\label{table: com}
\end{table}

Table \ref{table: com} shows the results for cross-lingual emotion identification using latent code learned by the proposed model and eGeMAPS. SVM trained on eGeMAPS shows the performance degradation of SER in cross-lingual scenarios compared to baseline results (Table \ref{table: baseline}) obtained using training and test data from the same corpus. Here, we use Urdu data as target language with no labels. Latent codes for source and target language data learned by the proposed model are given to the SVM. The source language data is used only for training and validation purposes. The testing is performed on target language data. The same evaluation is performed for other languages when Urdu data is used as source language (i.e., training data).  

\begin{table}[!ht]
\centering
\caption{UAR (\%) comparison using multi-lingual training.}
\begin{tabular}{|c|c|c|c|}
\hline
\multirow{2}{*}{Target Data} & \multicolumn{3}{c|}{UAR (\%)}                     \\ \cline{2-4} 
                             & eGeMAPS     & Latent Codes     & Latent Codes+eGeMAPS \\ \hline
EMO-DB                       & 60.5            & 65.9                & 68.0                    \\ \hline
SAVEE                        &  50.6           &  56.3             &  56.7                   \\ \hline
EMOVO                        &   56.8          &  60.5              & 61.8                    \\ \hline
URDU                         &  60.9      &   65.2              &     67.3               \\ \hline
\end{tabular}
\label{table: multi}
\end{table}
We also evaluated the proposed model using multi-language training as it helps to achieve better accuracy. Results are presented in Table \ref{table: multi}. In this experiment, we use the one-language-data-out scheme, and the remaining corpora are mixed and used for training the model. The results are compared using eGeMAPS and latent code learned by the proposed architecture of domain adaptation in Table \ref{table: multi}.  


\section{Summary of Findings}
\begin{figure}[!ht]
\centering
\includegraphics[trim=0.5cm 0cm 0.2cm 0.2cm,clip=true,width=0.5\textwidth]{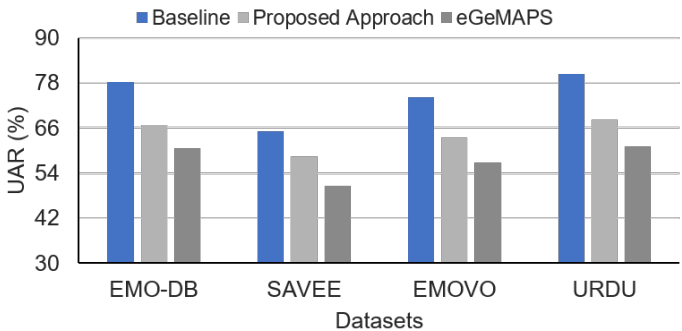}
\caption{UAR (\%) comparison of baseline (within corpus) results with multi-language training using eGeMAPS and proposed approach (latent codes).}
\label{fig:multi}
\end{figure}

We have presented results for cross-lingual SER using unsupervised adversarial domain adaptation. We were able to improve the results of cross-lingual SER significantly. From the experiments, best results are obtained when data from multiple languages are used as source data and one language as the target. This essentially means that the proposed model is capable of learning many intrinsic features from a broad range of languages which are similar to the target language. Table \ref{table: multi} shows that the results using the features learned by proposed model are significantly better compared to SVM trained on eGeMAPs using a similar multi-languages traning technique.  We also compared the results for multi-language training using the latent codes and eGeMAPS with baseline results. Fig. \ref{fig:multi} shows that baseline results using training and testing data from same corpus is better than latent codes learned by proposed model and eGeMAPS results in multi-lingual training scenario. However, the results using the proposed model outperformed significantly over the eGeMAPS. This shows the language invariant representations is able to improve the performance of cross-lingual emotion recognition. 
Overall, the results for cross-lingual emotion recognition have improved for all datasets without any requirement of labels for the target language dataset. This is a substantial leap towards better SER models. The findings of this work are of vital importance to many real-life SER applications where we have multiple resource data available. 

We also showed in Table \ref{table: com} that the performance of the existing SER systems degrade significantly across languages compared to the baseline results (Table \ref{table: baseline}) for all datasets. We have also presented results using the proposed model in Table \ref{table: com} that shows significant improvement in cross-lingual SER. More importantly, results using language invariant latent codes learned by the proposed model is even better than the multi-language training of SVM with eGeMAPS. For instance, we have achieved $64.25\%$ accuracy for Urdu data while using the EMO-DB as training data. This is better than $60.98\%$ that we have achieved using eGeMAPS by training SVM in multi-language training scheme with the remaining three datasets. Similar results are obtained for other datasets when they were used as target language data. This highlights the critical functionality of GAN based proposed architecture that can learn language invariant features in a completely unsupervised manner and provides improved results compared to directly using eGeMAPS. This shows that the proposed solution for cross-language SER can fit in scenarios where labelled data is not available for languages like Urdu. 

Another important insight learnt from this work is that the adversarially learned language invariant features when jointly used with eGeMAPS can help to achieve more excellent performance. The performance improvement is found for both cases including cross-lingual and multi-language training experiments. Results for cross-lingual SER using a combination of adversarially learned features (latent codes) and eGeMAPS are presented in Table \ref{table: com} and for multi-language training are presented in  Table \ref{table: multi}. 
\section{Conclusions}

In this paper, we proposed an unsupervised adversarial domain adaption approach for developing deep learning models for cross-lingual speech emotion recognition tasks. The proposed model is evaluated using the data from four emotional corpora. It is revealed that using GAN for learning language invariant features can provide better results compared to widely used emotional features like eGeMAPS. The proposed approach works in a completely unsupervised way, and adversarially learns language invariant features without the need of labels for the target language.



\end{document}